\documentstyle[12pt,epsfig]{article}
\begin{document}
\begin{center}
{\bf{ \boldmath
SEARCH FOR RARE $\phi$ DECAYS IN $\pi^+\pi^-\gamma$ FINAL STATE
}}
\end{center}
\begin{center}
R.R.Akhmetshin, G.A.Aksenov, E.V.Anashkin, M.Arpagaus, V.A.Astakhov,
B.O.Baibusinov, V.S.Banzarov, L.M.Barkov, A.E.Bondar,
D.V.Chernyak, S.I.Eidelman, G.V.Fedotovich, N.I.Gabyshev,
A.A.Grebeniuk, D.N.Grigoriev, P.M.Ivanov, B.I.Khazin, I.A.Koop, 
L.M.Kurdadze, A.S.Kuzmin,
I.B.Logashenko, P.A.Lukin, A.P.Lysenko, A.V.Maksimov, Yu.I.Merzlyakov, 
I.N.Nesterenko, V.S.Okhapkin, E.A.Perevedentsev, E.V.Popkov,
T.A.Purlatz, N.I.Root, A.A.Ruban, N.M.Ryskulov,
M.A.Shubin, B.A.Shwartz, V.A.Sidorov,
A.N.Skrinsky, V.P.Smakhtin, I.G.Snopkov, E.P.Solodov, A.I.Sukhanov,
V.M.Titov, Yu.V.Yudin.
\\
  Budker Institute of Nuclear Physics, Novosibirsk, 630090, Russia
\end{center}
\begin{center}
           D.H.Brown, B.L.Roberts
\\
                   Boston University, Boston, MA 02215, USA
\end{center}
\begin{center}
                J.A.Thompson, C.M.Valine
\\
                University of Pittsburgh, Pittsburgh, PA 15260, USA
\end{center}
\begin{center}
             V.W.Hughes
\\
                    Yale University, New Haven, CT 06511, USA 
\end{center}

%
%
\vspace{0.7cm}
\begin{abstract}
\hspace*{\parindent}
A search for $\phi$ radiative decays has been performed
using a data sample of about 2.0 million $\phi$ decays
collected by the CMD-2 detector at VEPP-2M collider in Novosibirsk.
From the selected $e^+e^-\rightarrow\pi^+\pi^-\gamma$ 
events the following results were obtained:

B($\phi\rightarrow f_{0}(980)\gamma) < 1\times10^{-4}$ for destructive
and 

B($\phi\rightarrow f_{0}(980)\gamma) < 7\times10^{-4}$ for 
constructive interference with the \\
Bremsstrahlung process respectively, 

B($\phi\rightarrow\gamma\rightarrow\pi^+\pi^-\gamma) < 3\times10^{-5}$ 
for $E_{\gamma}>20$ MeV,

B($\phi\rightarrow\rho\gamma) < 7\times10^{-4}$. 
\\
From the selected  $e^+e^-\rightarrow\mu^+\mu^-\gamma$ events 
 
 B($\phi\rightarrow\mu^+\mu^-\gamma) =
(2.3\pm1.0)\times10^{-5}$ has been obtained for $E_{\gamma}>20$ MeV.
\\
The upper limit on the P,CP-violating decay $\eta\rightarrow\pi^+\pi^-$ 
has also been placed:
  
B($\eta\rightarrow\pi^+\pi^-) < 9\times10^{-4}$ .  \\
All upper limits are at 90\% C.L.
\end{abstract}
\baselineskip=17pt
\section*{ \boldmath Introduction}
\hspace*{\parindent}
The radiative decays of the $\phi$ meson are accessible  
 at the VEPP-2M collider\cite{vepp2m} in Novosibirsk as well as at
TJNAF (former CEBAF) \cite{cebaf} 
and the future Novosibirsk and Frascati $\phi$-factories
\cite{skrinsky91,vignola92}. These decays were studied before 
(for a review see \cite{dol91}) and 
are now under study with SND\cite{SND}
and CMD-2\cite{CMD285,cmd2gen} detectors at VEPP-2M.
 
The decay $\phi\rightarrow f_{0}(980)\gamma$ 
is particularly interesting 
in two ways:

1) $f_{0}(980)$ structure.
The 20$\%$ decay probability into  a  two
kaon final state \cite{pdg} seems puzzlingly high
if $f_{0}(980)$ is a member of the strangeness-0 scalar meson
nonet.  Various explanations for this large  coupling  to  kaons  have
been advanced
\cite{acha,somef0,close91,f0theor}, including the
idea that $f_{0}(980)$ is composed  of  four
quarks, with a "hidden strangeness" component:
$
 (f_{0} = s~ \overline s ( u~ \overline u~+~ d~\overline d)/\sqrt2) ,
$
  or  that  it  may   be   a   $K-\overline{K}$
molecule.  The value of the branching ratio for the
$\phi\rightarrow f_{0}(980) \gamma$ decay mode 
appears to be very sensitive to the model \cite{acha,acha97}. 

2) Possible background to the planned 
measurement of $\epsilon'/\epsilon$ at $\phi$-factories.
The presence in the final state only the $K_S K_L$ from the $\phi$
decay is of crucial
importance for these experiments. The $\phi$ decay into $f_{0}(980) \gamma$ 
accompanied
with a low energy photon escaping detection leads to the C-even
final state of $K_S K_S$ and can mimic the CP-violating decay. 
In accordance with \cite{coccolino,franzini91}
this effect becomes significant if the final
state has a C-even component as large as $5\times10^{-5}$. Although 
theoretical predictions for the probability
of this decay give the values about $10^{-6}$ or even much less 
\cite{acha,somef0,close91,rosner}, 
the experimental measurement is needed for their confirmation.

The decay $\phi\rightarrow f_{0}(980)\gamma$ with $f_{0}(980)$ decaying to
two kaons is expected to be too small to be observed 
with the VEPP-2M collider luminosity
because of additional suppression caused by the  
small differences between the $\phi$ mass and kaon pair production threshold.
But it can be probed through the $f_{0}(980)$ decay
to two charged \cite {est90,eid91,fran92} or two neutral 
\cite{fran93} pions. In \cite{acha97} the branching ratio of the decay 
$\phi\rightarrow f_{0}(980)\gamma\rightarrow\pi\pi\gamma$ 
is predicted at the level of $(1-2)\times10^{-5}$ in the
$K-\bar{K}$ molecule model and $5\times10^{-5}$ for the conventional two quark
structure while it is $2.4\times10^{-4}$ in the four-quark model
\cite{acha}. Thus, observation
or even an upper limit for this decay mode will help to distinguish
between different possible structures of the $f_{0}(980)$ meson. 

In this paper we present results of a search for the 
$\phi\rightarrow f_{0}(980)\gamma$ decay in the event sample 
where two charged pions 
and one photon were detected in the CMD-2 detector.
The events in our sample arise primarily from the
much larger background: the radiative processes 
$e^+e^-\rightarrow\pi^+\pi^-\gamma$ where the photon comes from initial 
electrons or from final pions. Therefore the signal from the 
$f_{0}(980)\gamma$ final state can be seen most effectively 
as an interference structure 
at $E_{\gamma} \approx $ 40 MeV in the spectrum of the photons
and depends on the $f_{0}$(980) mass (980$\pm$10 MeV) 
and width (40-100 MeV) \cite{pdg}. 
\begin{figure}[tbh]
\begin{center}
\mbox{\epsfig{figure=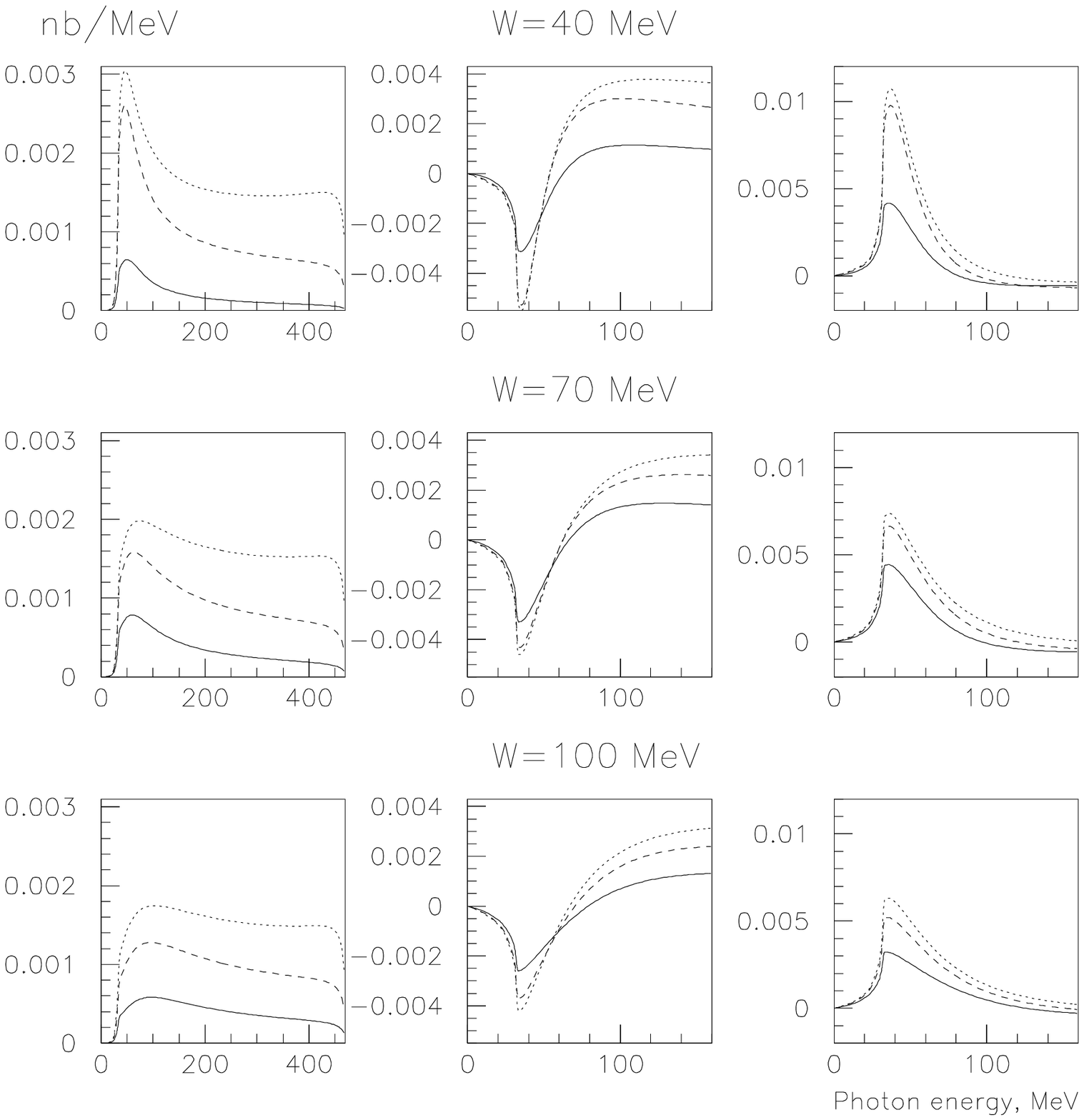,width=1.0\textwidth}}\\
\caption{
The signal from $\phi\rightarrow f_{0}(980)\gamma$ decay in the photon
spectra in the case of the four quark model for the $f_{0}(980)$ width of
40, 70 and 100 MeV and three values of branching ratios - 
$1,~3,~5\times10^{-4}$ (solid, dashed and dotted lines respectively). 
The first column is for the $f_{0}$ signal only, the second and third are 
the sum of the $f_{0}$ signal and interference
of $f_{0}$ with the radiative process $e^+e^-\rightarrow\pi^+\pi^-\gamma$ 
with positive and negative relative phase. The main contribution from 
$e^+e^-\rightarrow\pi^+\pi^-\gamma$ is not shown.
}
\label{f0-40-70-100}
\end{center}
\end{figure}
Figure \ref{f0-40-70-100} presents the photon energy spectra calculated
according to the four quark model \cite{acha97,acha96} for three 
values of the $f_{0}(980)$ width  and three values of the branching 
ratio. 
Two parameters of this model were varied
to keep the $f_{0}(980)$ width 40, 70 and 100 MeV. 
The $f_{0}(980)$ mass was fixed at 980 MeV.
The first column presents a signal for 
$f_{0}(980)$ only, while the second and third are the sum of the 
$f_{0}(980)$ signal 
and its interference with the radiative process 
$e^+e^-\rightarrow\pi^+\pi^-\gamma$ 
for positive and negative relative phases.
The main contribution from $e^+e^-\rightarrow\pi^+\pi^-\gamma$ 
(0.03 nb/MeV for 20 MeV photons, falling to
0.015 nb/MeV for 160 MeV photons) 
is 3 to 10 times higher 
than the maximum interference signal and is not shown. 

The signal was also searched as an interference structure in 
the energy dependence of the 
$e^+e^- \rightarrow \pi^+\pi^-\gamma$ cross section. 
\section*{ \boldmath The CMD-2 Detector}
\hspace*{\parindent}
The  CMD-2  detector  has  been
described  in  detail  elsewhere \cite{CMD285,cmd2gen}.

The CsI barrel calorimeter with 6$\times$6$\times$15 cm$^{3}$ crystal size
covering polar angles from 0.8 to 2.3 radian (892 crystals) 
is placed outside
a 0.4 r.l. superconducting solenoid with a 1 Tesla azimuthally symmetric
magnetic field.  The endcap calorimeter is made of 680
BGO crystals with 
2.5$\times$2.5$\times$15 cm$^{3}$ size
and was not installed for the data presented here.
The   drift chamber (DC) inside the solenoid  has 
the momentum resolution of 6$\%$ for 500 MeV/c  charged particles.
The energy resolution for photons in the CsI calorimeter is about 8$\%$.
The muon  system uses  streamer tubes grouped in two layers (inner and
outer) with a 15 cm magnet yoke serving as an absorber and has 1-3 cm spatial 
resolution.

 The integrated luminosity of 1480 $nb^{-1}$ collected in 1993 
in 14 energy points around the $\phi$ mass corresponds to
the production of about $2.0 \times 10^{6} \phi$ 's.
About
$7.2 \times 10^{7}$ events, predominantly beam-beam, beam-gas and Bhabha
events were recorded. 
Some preliminary results from this data set 
have been published in \cite{cmdpre93}.

%
%
%
%
%
\section*{ \boldmath Selection of $\pi^+\pi^-\gamma$ Events} 
\hspace*{\parindent}
The event candidates were selected by a  requirement of  only  two
(minimum ionizing)
charged tracks in the DC 
and one or two photons with energy greater than 20 MeV  in
the CsI calorimeter.  Events with invariant mass of two photons close to the
$\pi^0$ mass were removed. 
The average momentum of two charged tracks was required 
to be higher than 240 MeV/c to remove the background from 
$K_{S} \rightarrow \pi^{+}\pi^{-}$ decays and the radial distance 
of the closest approach of each track to the interaction region was 
required to be less than 0.3 cm.
The requirement that the  Z-coordinate (along the beam) of the vertex be
within 10 cm at the detector center reduces cosmic ray 
background by a factor of two.

The detected photons were required to have a polar angle between
0.85 and 2.25 radians so that they enter
the "good" region in the CsI barrel calorimeter.

The detected charged tracks were required to have a polar angle between
1.05 and 2.1 radians so that they enter the inner muon system.
The sum of the energy depositions of two clusters 
associated with two charged tracks is required to be less than 400 MeV. 
These cuts removed Bhabha events.

The main  background for the $\pi^{+}\pi^{-}\gamma$ final state 
after the above cuts comes from:
a) the radiative process $e^+ e^- \rightarrow\mu^+\mu^-\gamma$,
b)  the decay $\phi \rightarrow \pi^{+}\pi^{-}\pi^{0}$ when one of the photons
from the $\pi^0$ escapes detection, and
c) collinear events $e^{+}e^{-} \rightarrow \mu^{+}\mu^{-},\pi^{+}\pi^{-}$,
with a background cluster in the calorimeter from secondary decays
and interactions of muons or pions with the detector material.
\section*{ \boldmath Selection of Muons}
\hspace*{\parindent}
The inner muon  system was used to separate muons from pions.
The requirement of hits 
in the inner muon system for both charged tracks selects muon events,
together with some pion events in which the products of the nuclear 
interactions in the calorimeter reach the muon system.

\begin{figure}[tbh]
\mbox{\epsfig{figure=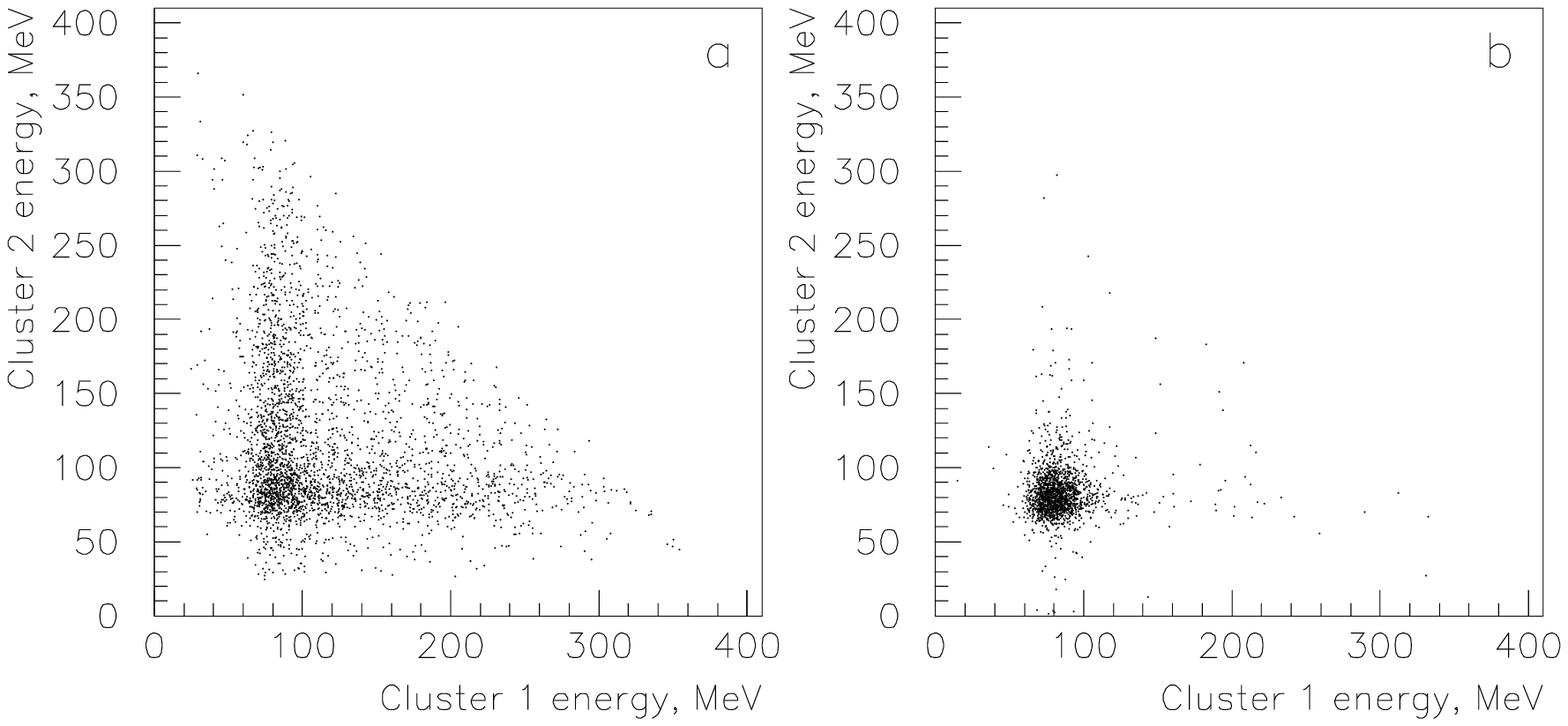,width=1.0\textwidth}}\\
\vspace{-7.5cm}
\caption{Calorimeter response for events with 1 or no hits in the muon system 
(a) and for events selected as muons (b).
}
\label{ec1-ec2}
\end{figure}

Separation of pion and muon events in the calorimeter
is illustrated in Fig.~\ref{ec1-ec2}, where
scatter plots of energy depositions of one track vs. the other one
are presented for events with one or no hits in the muon system (a) and
selected as muons (b). Energy depositions are
corrected for the incident angle.  Pions have nuclear interactions and 
in some cases leave more energy, while muons mostly have only dE/dx losses.

%
%

To select a cleaner sample of  muon events,
in  addition to the information from the muon system 
both tracks were required to show only minimum ionizing energy deposition
in the calorimeter (60-130 MeV).
All the rest were considered as candidates to pion sample.


The average efficiency of the muon system with above selections 
was estimated to be (91$\pm$3)\%. The systematic error is estimated to be 
about 5\% and results in a correlated uncertainty in the selected 
number of pions and muons.
%
%

%

Simulation of $\pi^{+}\pi^{-}\gamma$ events shows that the probability
for pions to be selected as muons is about 20\% at photon energies 
below 150 MeV and falls to 5\% at 350 MeV photon energy. 
\section*{ \boldmath Constrained Fit}
\hspace*{\parindent}
To reduce the background from collinear events as well as that from three 
pions a constrained fit was used requiring total energy and momentum 
conservation (within detector resolution) for a three body decay. 
About 20\% of the selected events had an additional photon. In this case
the constrained fit was applied for both possible combinations and 
the combination with minimum $\chi^2$ was chosen.

\begin{figure}[tbh]
\mbox{\epsfig{figure=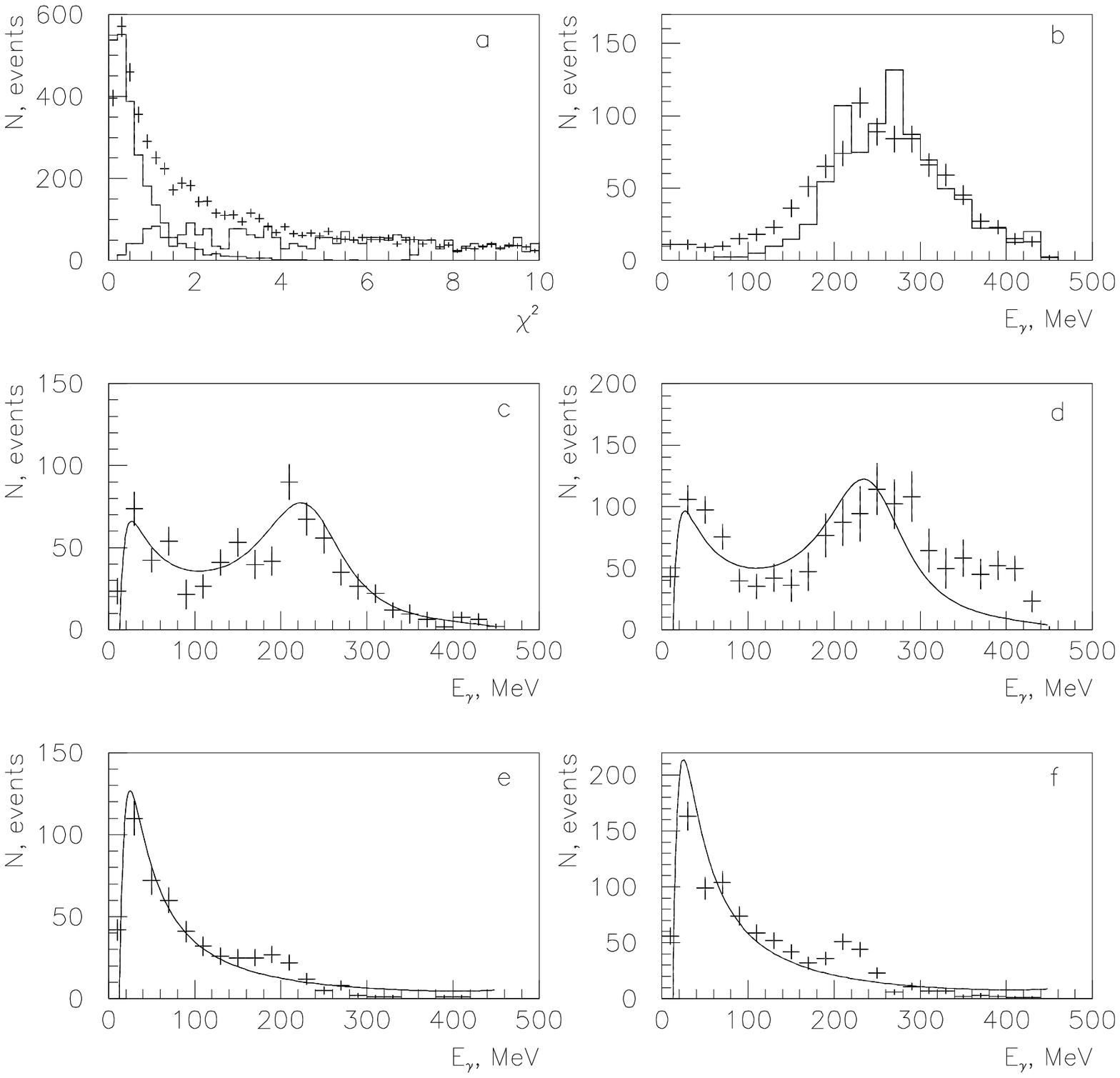,width=1.0\textwidth}}\\
\caption{
Event selection. a-$\chi^2$/d.f. distribution for events (dots with 
errors) and 
simulation(histograms); b-photon energy spectrum for background events
(histogram is for 3 pion simulation);
c,d-photon energy spectra for $\pi^+\pi^-\gamma$ events after 
background subtraction for the "non-$\phi$" region (c) 
and "$\phi$" region (d) (lines are calculation); 
e-f - the same photon energy spectra for $\mu^+\mu^-\gamma$ events.
}
\label{evsel}
\end{figure}

Figure \ref{evsel}a presents the $\chi^2$/d.f. distribution 
(points with errors) for selected
$\pi^+\pi^-\gamma$ events after the above 
cuts and the constrained fit. The peaked at zero distribution for simulated 
events of interest is shown as a histogram. The experimental distribution has
a "tail" due to the above background processes. This tail is mostly 
dominated by the
three pion events as demonstrated by a flat
simulated histogram. 
A three pion background appears when one of the photons 
from the $\pi^{0}$ has energy below 20 MeV and is not detected
so that the event looks like a three body decay within detector resolution.

The $\chi^2$/d.f. distribution for events selected as muons has much 
less background and is in good agreement with simulation. 
A cut on the $\chi^2$/d.f. less than 3 was imposed.
 
To obtain the correct number of
events and spectra the background subtraction procedure was used. 
The events with 3 $<~\chi^2$/d.f.$~<$ 6 were selected to obtain the
background spectrum which is shown in
Figure \ref{evsel}b (points with errors).
The histogram shows the photon spectrum 
for simulated three pion events with the same cuts. 
The low energy part of the background spectrum has a big
contribution from collinear events. For each energy point 
the background was subtracted from the events in the region
$\chi^2$/d.f.$~<~$ 3 used in the further analysis. 

As a result of the constrained fit one can
obtain an improved estimate for the photon energy. 
Simulation shows a photon  energy resolution  about 5 MeV in the whole 
energy range instead of $\sigma_{E_{\gamma}}=8\%\times E_{\gamma}$ 
CsI resolution. 

%
%

In order to extract the resonant contribution associated with the
$\phi$, two data sets were used. Energy points
at E$_{c.m.}$ from 1016 to 1023.2 MeV with the integrated luminosity of
$897~nb^{-1}$ were used for the "$\phi$" region. 
The points at
E$_{c.m.}$ = 996, 1013, 1026, 1030, 1040 MeV with the integrated luminosity
of $584~nb^{-1}$ were used for a background estimate 
from a "non-$\phi$" region, containing 30 times fewer $\phi$ decays
than  the "$\phi$" region.

Figures \ref{evsel}c,d present photon spectra corrected by the constrained fit
after background subtraction
for events with $\chi^2/d.f.~<$3 for "non-$\phi$" (c) and $\phi$ (d)
regions.
Lines present calculated spectra corrected for the  detector efficiency
(see below).

A peak at 220 MeV corresponds to the radiative process 
$e^+e^-\rightarrow\rho\gamma$ with the $\rho$ decay into two charged pions.
Figure \ref{evsel}d also shows the presence of some background photons
with energies higher than 300 MeV presumably coming from the $\phi$ decay into 
$\eta\gamma$ with $\eta\rightarrow\pi^+\pi^-\gamma$. If a soft photon
from $\eta$ is not detected, the constrained fit changes the energy of
the remaining  photon from its
362 MeV peak value within the detector resolution but the $\chi^2$/d.f. value
remains good. These events are
not completely removed by the subtraction procedure. 

The region with the photon energy from 20 to 160 MeV has minimum background 
and has been chosen for the $\phi\rightarrow f_{0}(980)\gamma$ search.

Figures~ \ref{evsel}e,f present the same distributions for events
selected as muons. Lines are calculated spectra corrected for the detection
efficiency.
The presence of pion events in the muon sample is illustrated 
by the small peak at 220 MeV from  $\rho$ decays.

\section*{ \boldmath Cross Section Calculation}
\hspace*{\parindent}
The cross section for each energy point was calculated as $\sigma$ = 
$N_{ev}$/(L$\cdot\epsilon$).

$N_{ev}$ is the number of selected events with photons  
having polar angles from 0.85 to
2.25 radians and energy from 20 to 160 MeV and charged particles in the
polar angle range 1.05-2.1 radian. 
The final numbers of muons and pions were corrected
for their cross contamination.

The integrated luminosity L for each energy point was determined by Bhabha
events with about 2\% systematic accuracy.

The detection efficiency $\epsilon$ was obtained by simulation. 
The CsI single photon efficiency is about 80\% for 20 MeV photons and
reaches 100\% level above 70 MeV.
The $\pi^+\pi^-\gamma$ and $\mu^+\mu^-\gamma$ detection efficiency decreases 
with photon energy increase because the acceptance is higher for nearly 
collinear tracks and is on the average 20\%. 
%
%
%
The center of mass energy, integrated luminosity for each
energy point, the  number of selected events with pions and muons 
and corresponding cross sections 
are presented in Table 1. 
The errors of the cross sections
are statistical only.  
The systematic error in the experimental cross sections was estimated to
be about 6\% dominated by the uncertainty of the muon system efficiency.
\section*{ \boldmath Determination of the Branching Ratios}
\begin{figure}[tbh]
\mbox{\epsfig{figure=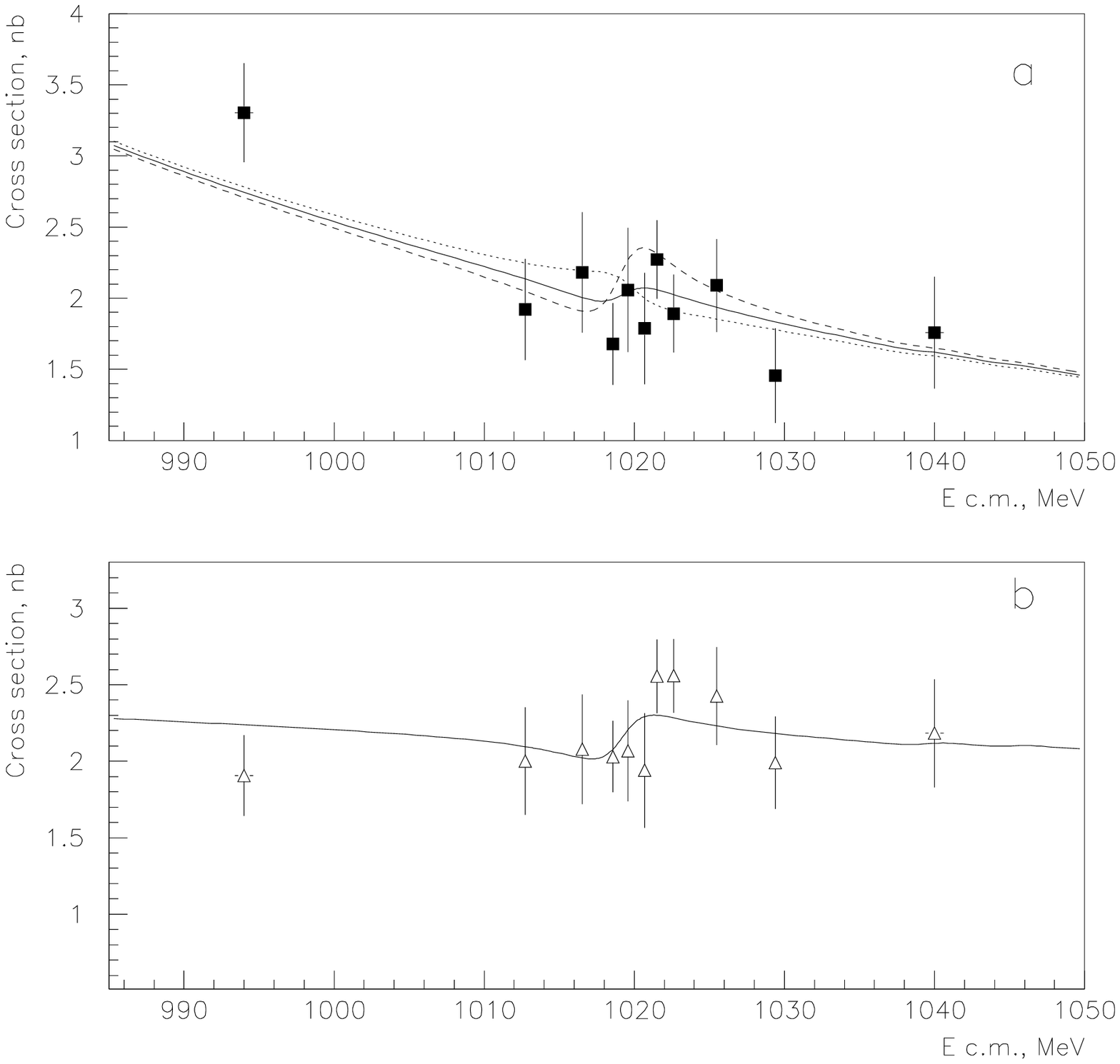,width=1.0\textwidth}}\\
\caption{
$\phi\rightarrow f_{0}(980)\gamma$ search:
a. The cross section for 
$e^{+}e^{-}\rightarrow\pi^{+}\pi^{-}\gamma$. Lines are theoretical
predictions in case of no $f_{0}(980)\gamma$ signal (solid line), 
$f_{0}(980)\gamma$ 
signal with B($\phi\rightarrow f_{0}\gamma$)=2.4$\times10^{-4}$
for the positive (dashed line) and negative relative 
phase (dotted line); 
b. The cross section for $e^{+}e^{-}\rightarrow\mu^{+}\mu^{-}\gamma$
with the theoretical prediction.
}
\label{fnot-gamma}
\end{figure}

\hspace*{\parindent}
The obtained cross sections of the processes
$e^+e^-\rightarrow\pi^+\pi^-\gamma$ and
$e^+e^-\rightarrow\mu^+\mu^-\gamma$  versus energy
are presented in Fig. \ref{fnot-gamma}a,b. 
%
To interpret the  data we follow the model described  in \cite{acha96}.
The resulting cross section (the formula for its energy dependence is 
not presented here because of its cumbersomeness) contains contributions 
from different reactions: Bremsstrahlung by initial and final
particles as well as the direct hadronic decay 
$\phi\rightarrow \pi^{+}\pi^{-}\gamma$.
 
The process of Bremsstrahlung from initial particles is suppressed by 
selecting photons transverse to the beam 
direction, but it still accounts for 2/3 of the  
$e^+e^-\rightarrow\pi^+\pi^-\gamma$ 
and one half of  the $e^+e^-\rightarrow\mu^+\mu^-\gamma$ cross section.

Also taken into account is the $\phi$ contribution to the photon
propagator (vacuum polarization). This 
contribution gives rise to an interference pattern  in the cross section 
at the $\phi$ mass and can be referred to as a "radiative" decay 
$\phi\rightarrow\gamma\rightarrow\pi^+\pi^-\gamma$.
The amplitude of the interference is determined by the  
$\phi$-meson leptonic width. 
%
%
The same is valid for the $\mu^+\mu^-\gamma$ final state.
 

The direct hadronic decay 
$\phi\rightarrow\pi^+\pi^-\gamma$ is expected to be dominated
by the  $\phi\rightarrow f_{0}(980)\gamma$ decay.
If the amplitude of this process is high enough, the interference pattern
in the cross section is predicted to be different from that with the vacuum 
polarization contamination only.  

To fit our data the calculations performed in 
\cite{acha96} were used in the case
when $f_{0}(980)$ is a four quark state. 
The four quark model predicts for the
branching ratio B($\phi\rightarrow f_{0}(980)\gamma$)=2.4$\times10^{-4}$, 
close to the sensitivity of our experiment while
the two quark and kaon molecule models predict values below our
sensitivity and were not used for fitting. 
 
Using the above descriptions the following results have been obtained:
  
1. If direct hadronic decays of the $\phi$ give no contribution
to the observed $\pi\pi\gamma$ events,
the fit with only hadronic vacuum polarization pattern gives:
\\

B($\phi\rightarrow\pi^+\pi^-\gamma)<3.0\times 10^{-5}$ at 90\% CL,

B($\phi\rightarrow\mu^+\mu^-\gamma$)=(2.3$\pm$1.0$)\times10^{-5}$.
\\


The present upper limit for the first decay is
B($\phi\rightarrow\pi^+\pi^-\gamma)<7\times 10^{-3}$ \cite{cosme} while
there are no other measurements for the 
B($\phi\rightarrow\mu^+\mu^-\gamma$). 
These results correspond to the photon energy E$_{\gamma}>~$20 MeV
and don't contradict to the
theoretical calculations for these processes \cite{acha96}:
\\

B($\phi\rightarrow\pi^+\pi^-\gamma$)=4.74$\times10^{-6}$,

B($\phi\rightarrow\mu^+\mu^-\gamma$)=1.15$\times10^{-5}$.
\\


For the experimental 
range of angles and photon energies, the ratio $\sigma_{exp}/\sigma_{th}$ 
for pions and muons
is 0.95$\pm$0.05 and 0.98$\pm$0.04 respectively showing good agreement 
with theory. 
Theoretical predictions with the $\phi$-meson leptonic width from \cite{pdg}
are shown 
in Fig.~\ref{fnot-gamma}a (solid line) and  
Fig.~\ref{fnot-gamma}b, 
the $\chi^2$/d.f. are 7.4/10 and 4.8/10 respectively.

2. Taking the value of the leptonic width from \cite{pdg}, we are fixing  
the vacuum polarization contribution to the decay
$\phi\rightarrow\pi^+\pi^-\gamma$. After that  
the following upper limits for $f_{0}(980)\gamma$ 
production have been obtained from the c.m. energy dependence 
of the cross section and 
two different signs of the interference at 90\% CL:
\\

B($\phi\rightarrow f_{0}(980)\gamma)<10\times10^{-4}$ for "+",

B($\phi\rightarrow f_{0}(980)\gamma)<4\times10^{-4}$ for "--".
\\

The results above correspond to all possible photon energies. 
The predicted cross sections in the case of the four quark model
are presented in Fig.~\ref{fnot-gamma}a for two signs of relative
phases.

\begin{figure}[tbh]
\mbox{\epsfig{figure=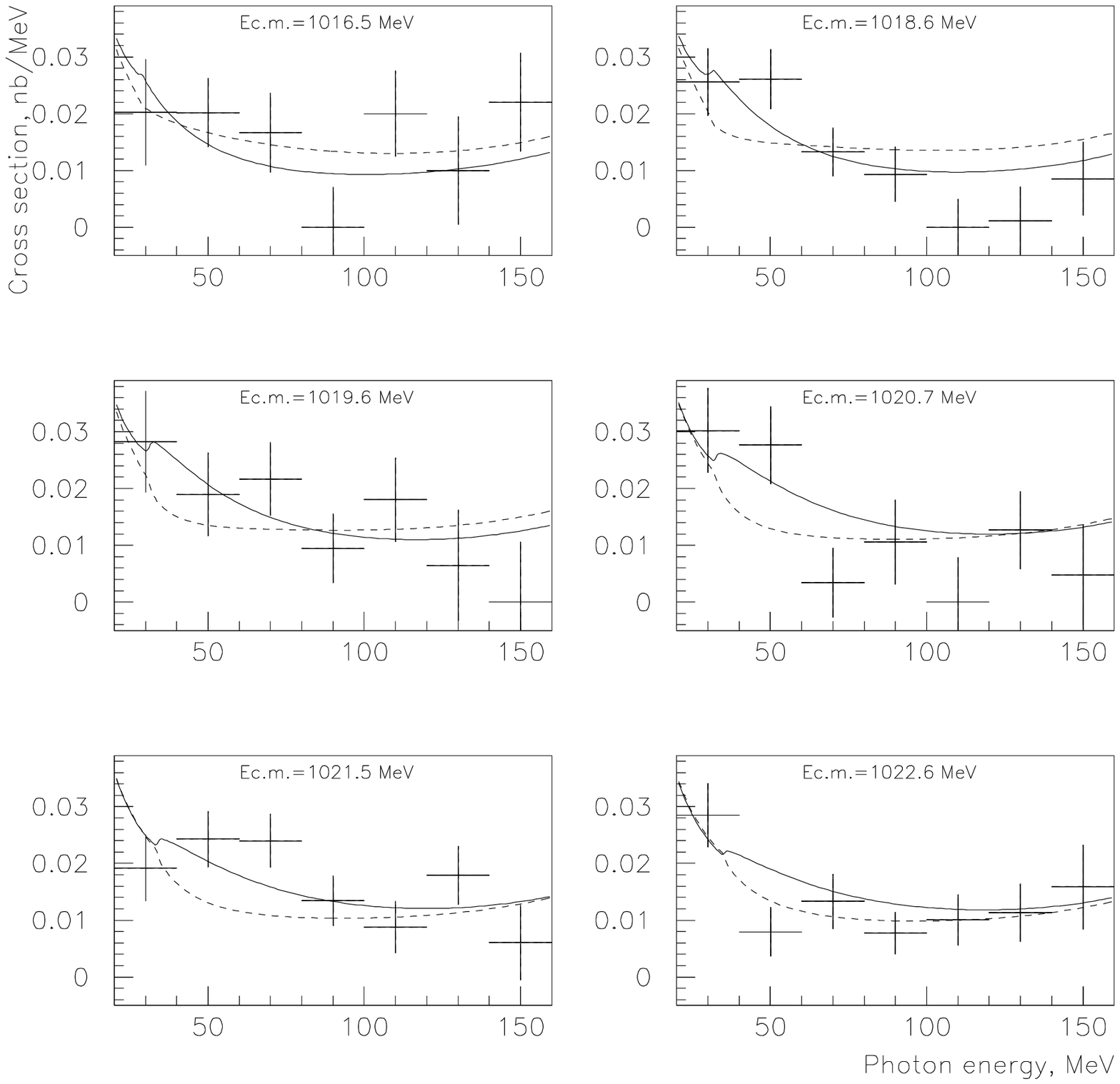,width=1.0\textwidth}}\\
\caption{$\phi\rightarrow f_{0}(980)\gamma$ search:
Photon spectra for six c.m.energy point in the "$\phi$" region.
Lines are theoretical predictions for the four quark model for the
branching ratio of 
$2.4\times 10^{-4}$ in the case of the positive (solid line) and
negative (dashed line) interference sign and a 70 MeV $f_{0}(980)$ width.
} 
\label{f0-spectr}
\end{figure}

 The signal from the decay of the $\phi \rightarrow f_{0}(980)\gamma$ would be
also seen as a structure in the radiative
differential photon spectra from the
"$\phi$" region. These spectra are  
shown in Fig.~\ref{f0-spectr} for six c.m.energy points 
for photons in the 20-160 MeV energy range. The spectra were corrected for all
experimental inefficiencies and normalized to the integrated luminosity.
Also shown is the theoretical prediction  from the four quark 
model \cite{acha96,acha97} for two possible signs of interference. 
%

The above spectra were fit as a group, with the photon spectrum 
calculated for each energy point taking into account the uncertainties
in the $f_{0}$(980) mass and width.  
As a result, lower upper limits have been obtained at $90\%$ CL:
\\

B($\phi\rightarrow f_{0}(980)\gamma)<7\times10^{-4}$ for "+",

B($\phi\rightarrow f_{0}(980)\gamma)<1\times10^{-4}$ for "--".
\\

%
%
\section*{ \boldmath Search for $\phi\rightarrow\rho\gamma$ Decay}
\begin{figure}[tbh]
\mbox{\epsfig{figure=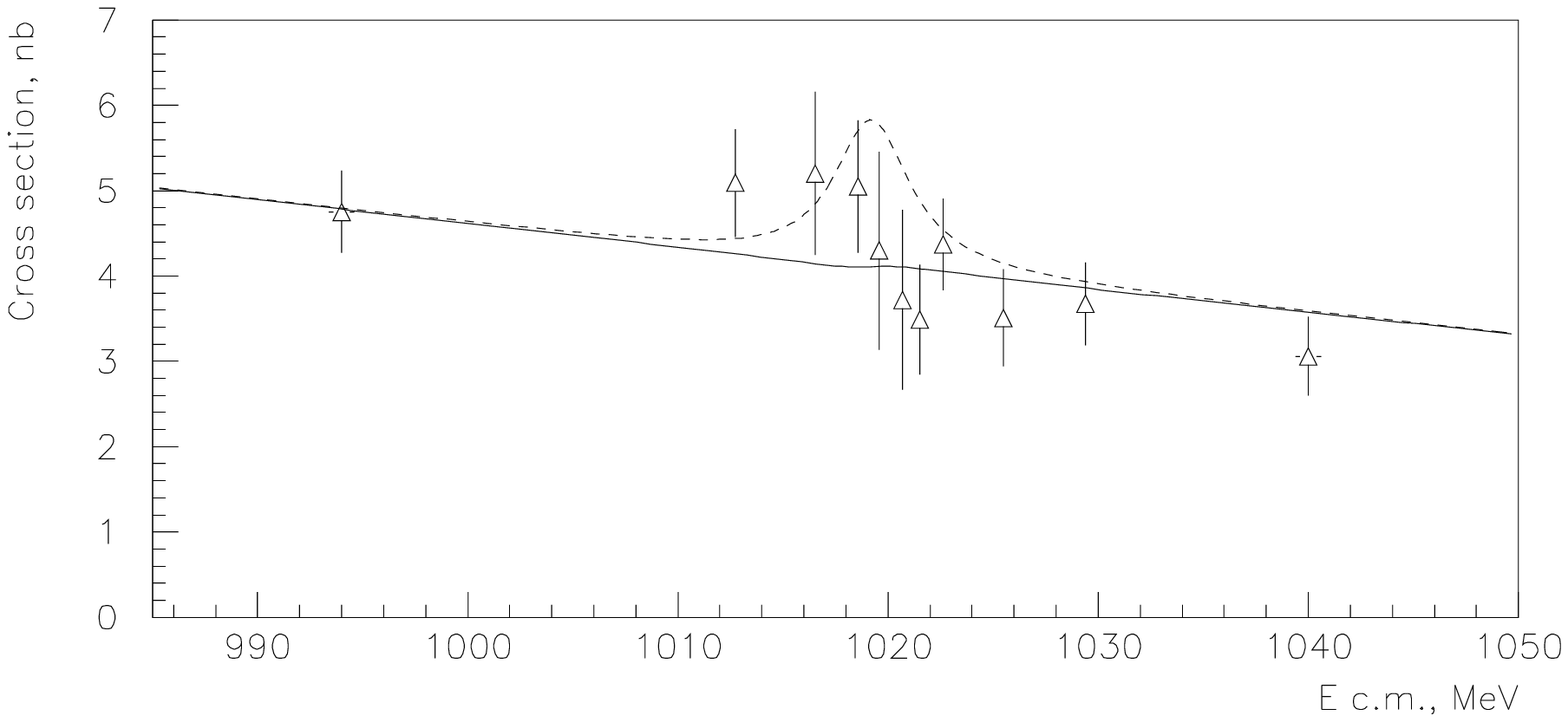,width=1.0\textwidth}}\\
\vspace{-7.5cm}
\caption{$\phi\rightarrow\rho\gamma$ search.
The cross section for $\pi^+\pi^-\gamma$ events for photons
in the 100-300 MeV region. The solid line is the theoretical prediction.
The dashed line shows the possible $\phi$ signal at 90\% C.L.
} 
\label{rho-sech}
\end{figure}

\hspace*{\parindent}
The selected $\pi^+\pi^-\gamma$ events with photon energies from
100 to 300 Mev (see Fig. \ref{evsel}d)
can be used to search for the C violating decay
$\phi\rightarrow\rho\gamma$ with the $\rho$ decay into two charged pions.
The cross section vs. energy for the events with photons in the 100-300 MeV
range is presented in Fig. \ref{rho-sech}. 
%
%
The fit of the energy dependence
of the cross section by the theoretical calculation \cite{acha96}
gives the ratio $\sigma_{exp}/\sigma_{th}$=1.04$\pm$0.05 and
shows that maximum
possible signal from the $\phi$ does not exceed 1.7 nb at 90\% C.L. 
It corresponds to the following upper limit:
\\

B($\phi\rightarrow\rho\gamma)< 7\times10^{-4}$ at 90\% C.L.
\\

which is thirty times better than the previous one
\cite{rhog}. 
\section*{ \boldmath Search for $\eta\rightarrow\pi^+\pi^-$ Decay}
\hspace*{\parindent}
The selected $\pi^+\pi^-\gamma$ events can be used to search for the
P and CP violating decay $\eta\rightarrow\pi^+\pi^-$, where the 
$\eta$ comes from
the radiative $\phi \rightarrow \eta\gamma$ decay. 
This decay should  be seen as a narrow  
peak at the photon energy of 362 MeV.

With 897 $nb^{-1}$
in the  "$\phi$" region 
the total number of produced $\eta$ was estimated as 26900 and can be 
used for normalization.

\begin{figure}[tbh]
\mbox{\epsfig{figure=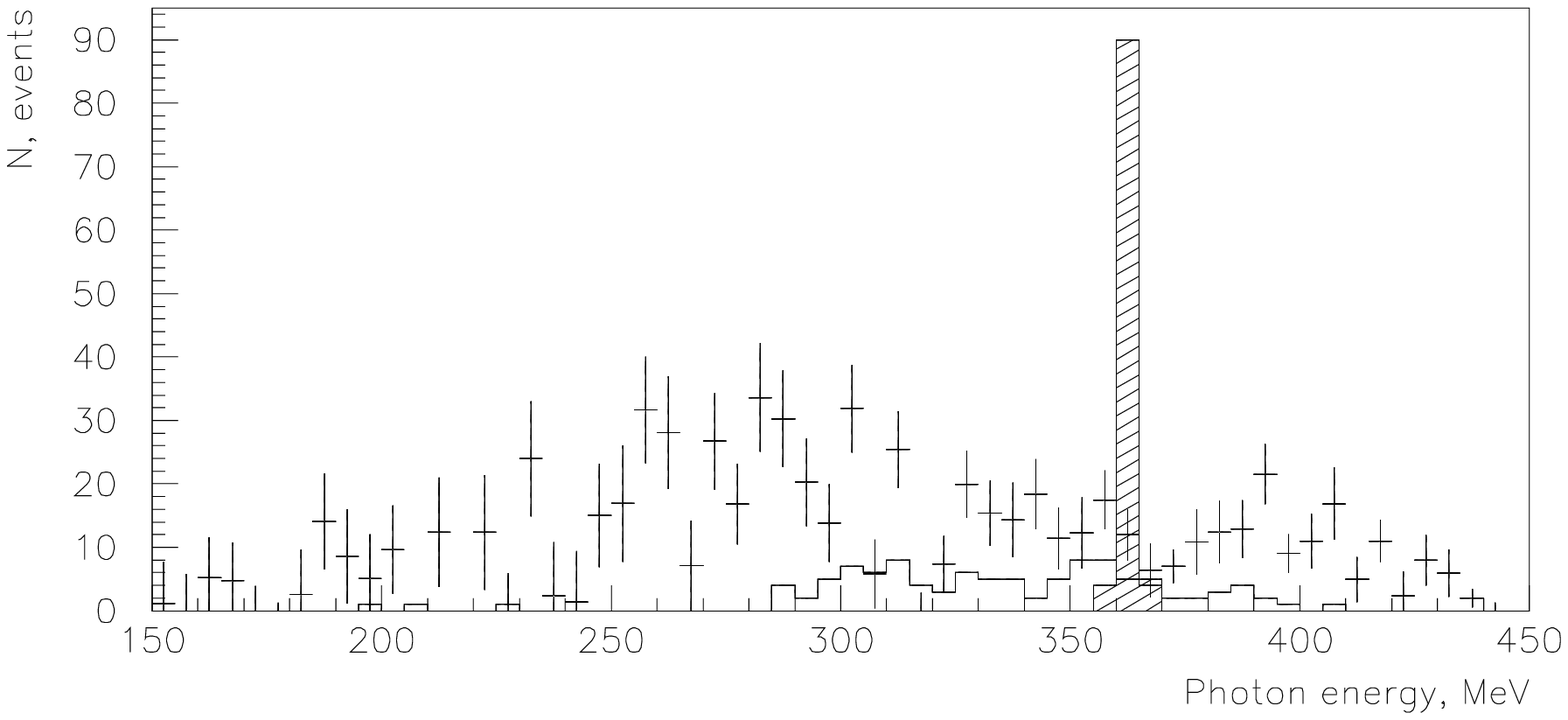,width=1.0\textwidth}}\\
\vspace{-7.5cm}
\caption{$\eta\rightarrow\pi^+\pi^-$ search.
The experimental photon spectrum for 
for $\pi^+\pi^-\gamma$ events from "$\phi$" region.
Histograms are simulation of the $\eta\rightarrow\pi^+\pi^-$ decay 
before and after (shaded) constrained fit.
} 
\label{eta_pipi}
\end{figure}

As mentioned above, the
constrained fit procedure improves the photon energy resolution to about 
5 MeV.  Figure \ref{eta_pipi} demonstrates this improvement where 
simulated photon spectra are shown by histogram before and after 
(shaded) constrained fit. The experimental spectrum for the "$\phi$" region is
shown by points with errors. 
Only background events are seen in the 362 MeV region and
the number of  possible $\eta\rightarrow\pi^+\pi^-$ 
events was estimated to be $N~<~4$.     

The detection efficiency for these events taken from simulation 
is (16$\pm$3)\%.  
The corresponding upper limit is:
\\

B($\eta \rightarrow\pi^{+}\pi^{-})~<~9\times 10^{-4}$ at 90$\%$ C.L.,
\\

which is a factor of 1.5 better than the only existing one from 
\cite{etapipi}.

%
%
%
\section*{ \boldmath Conclusions}
\hspace*{\parindent}
Using about 30\% of available data at the $\phi$-meson the selection of 
$e^+e^-\rightarrow\pi^+\pi^-\gamma$ and 
$e^+e^-\rightarrow\mu^+\mu^-\gamma$ events was performed. A new upper limit
for the $\phi\rightarrow f_{0}(980)\gamma$ decay has been obtained

B($\phi\rightarrow f_{0}(980)\gamma)~<~(1-7)\times10^{-4}$ at 90\% C.L.\\
depending on the sign of interference with the radiative process
$e^+e^-\rightarrow\pi^+\pi^-\gamma$. This result can be compared with 
the only existing B($\phi\rightarrow f_{0}(980)\gamma)<2\times10^{-3}$
\cite{dol91}.

Assuming that the contribution of the direct hadronic decay of $\phi$
into $\pi^+\pi^-\gamma$ is negligible, 
the following result has been obtained:

B($\phi\rightarrow\gamma\rightarrow\pi^+\pi^-\gamma)~<~3\times10^{-5}$ at 90\% 
C.L. and E$_{\gamma}>$20 MeV\\
which should be compared with 
B($\phi\rightarrow\pi^+\pi^-\gamma)<7\times10^{-3}$ obtained in \cite{cosme}.
\\
The result

B($\phi\rightarrow\gamma\rightarrow\mu^+\mu^-\gamma)=(2.3\pm1.0)\times10^{-5}$
\\
for photon energies $E_{\gamma}>20 MeV$ 
is obtained for the first time.

For the C violating decay of $\phi\rightarrow\rho\gamma$ and P,CP violating 
decay of $\eta\rightarrow\pi^+\pi^-$ the following results have been 
obtained:

B($\phi\rightarrow\rho\gamma)~<~7\times10^{-4}$ at 90\% C.L.,

B($\eta\rightarrow\pi^+\pi^-)~<~9\times10^{-4}$ at 90\% C.L.
\\
which are also better than previous ones \cite{rhog,etapipi}.

Analysis of the collected data 
is in progress and we expect new results on
the $\phi$ rare decays.
\subsection*{Acknowledgements}
\hspace*{\parindent}
The authors are grateful to N.N.Achasov and V.V.Gubin for useful
discussions and help with the data interpretation.

This work is supported in part by the US 
Departament of Energy, US
National Science Foundation and the
International Science Foundation under the grants RPT000 and RPT300.

M.Arpagaus acknowledges support from the Swiss National Science Foundation.
\newpage
\vspace {0.5cm}
\begin{center}
Table 1: Energy, Integrated luminosity, Number of Events and 
Experimental Cross Sections for $\pi^+\pi^-\gamma$ and 
$\mu^+\mu^-\gamma$ Channels.
\end{center}
\vspace {0.5cm}
\begin{tabular} {|c|c|c|c|c|c|}
\hline
 $E_{c.m.}$, MeV & L, $nb^{-1}$ 
& $N_{\pi^+\pi^-\gamma}$ 
& $N_{\mu^+\mu^-\gamma}$
& $\sigma_{\pi^{+}\pi^{-}\gamma}$, nb 
& $\sigma_{\mu^+\mu^-\gamma}$, nb\\ 
\hline
   994.000$\pm$0.600 & 152.0$\pm$0.6 & 122$\pm$13 &  74$\pm$10 
& 3.30$\pm$0.35 & 1.91$\pm$0.26 \\
  1012.734$\pm$0.076 & 110.9$\pm$0.5 &  52$\pm$ 9 &  76$\pm$10 
& 1.92$\pm$0.36 & 2.00$\pm$0.35 \\
  1016.546$\pm$0.068 &  85.5$\pm$0.5 &  48$\pm$ 9 &  44$\pm$ 8 
& 2.18$\pm$0.42 & 2.08$\pm$0.36 \\
  1018.585$\pm$0.340 & 204.3$\pm$0.7 &  87$\pm$14 & 103$\pm$12 
& 1.68$\pm$0.29 & 2.03$\pm$0.23 \\
  1019.566$\pm$0.062 & 102.0$\pm$0.5 &  45$\pm$10 &  48$\pm$ 8 
& 2.06$\pm$0.44 & 2.07$\pm$0.33 \\
  1020.676$\pm$0.062 &  89.1$\pm$0.5 &  37$\pm$ 9 &  43$\pm$12 
& 1.79$\pm$0.39 & 1.94$\pm$0.38 \\
  1021.519$\pm$0.121 & 203.2$\pm$0.6 & 117$\pm$14 & 128$\pm$13 
& 2.27$\pm$0.28 & 2.56$\pm$0.24 \\
  1022.638$\pm$0.372 & 212.4$\pm$0.6 &  99$\pm$14 & 136$\pm$10 
& 1.89$\pm$0.27 & 2.56$\pm$0.24 \\
  1025.586$\pm$0.070 & 117.9$\pm$0.5 &  61$\pm$ 9 &  72$\pm$10 
& 2.09$\pm$0.33 & 2.42$\pm$0.32 \\
  1029.402$\pm$0.064 & 121.7$\pm$0.6 &  40$\pm$ 9 &  55$\pm$ 8 
& 1.46$\pm$0.33 & 1.99$\pm$0.30 \\
  1040.000$\pm$0.600 &  81.8$\pm$0.5 &  35$\pm$ 8 &  45$\pm$ 7 
& 1.76$\pm$0.39 & 2.18$\pm$0.35 \\
\hline
\end{tabular}
\vspace {0.5cm}
\end{document}